\documentclass{emulateapj}

\begin{document}

\shortauthors{Todorov et al.}
\shorttitle{Planetary-mass Companion in Taurus}

\title{Discovery of a Planetary-mass Companion to a Brown Dwarf in 
Taurus\altaffilmark{1}}

\author{
K. Todorov\altaffilmark{2},
K. L. Luhman\altaffilmark{2,3},
and K. K. McLeod\altaffilmark{4}
}

\altaffiltext{1}{Based on observations performed with the NASA/ESA
{\it Hubble Space Telescope} and Gemini Observatory.
The {\it Hubble} observations are associated with proposal ID 11203 and
were obtained at the Space Telescope Science Institute, which is operated by
the Association of Universities for Research in Astronomy, Inc., under NASA
contract NAS 5-26555.}
\altaffiltext{2}{Department of Astronomy and Astrophysics, The Pennsylvania
State University, University Park, PA 16802; kot104@astro.psu.edu.}
\altaffiltext{3}{Center for Exoplanets and Habitable Worlds, The 
Pennsylvania State University, University Park, PA 16802.}
\altaffiltext{4}{Whitin Observatory, Wellesley College, Wellesley, MA.}

\begin{abstract}

We have performed a survey for substellar companions to young brown
dwarfs in the Taurus star-forming region using the Wide Field Planetary
Camera 2 on board the {\it Hubble Space Telescope}. In these data,
we have discovered a candidate companion at a projected separation of
$0\farcs105$ from one of the brown dwarfs, corresponding to 15~AU
at the distance of Taurus. To determine if this object is a companion,
we have obtained images of the pair at a second epoch with 
the adaptive optics system at Gemini Observatory.
The astrometry from the {\it Hubble} and Gemini data indicates that
the two objects share similar proper motions and thus are likely companions.
We estimate a mass of 5--10~$M_{\rm Jup}$ for the secondary based on
a comparison of its bolometric luminosity to the predictions of
theoretical evolutionary models.
This object demonstrates that planetary-mass companions to brown dwarfs
can form on a timescale of $\tau\lesssim1$~Myr.
Companion formation on such a rapid timescale is more likely to occur via
gravitational instability in a disk or fragmentation of a cloud core than
through core accretion. 
The Gemini images also reveal a possible substellar companion
($\rho=0\farcs23$) to a young low-mass star that is $12\farcs4$ from
the brown dwarf targeted by {\it Hubble}. If these four objects comprise a
quadruple system, then its hierarchical configuration would suggest
that the fragmentation of molecular cloud cores can produce companions
below 10~$M_{\rm Jup}$.

\end{abstract}

\keywords{
protoplanetary disks --- 
stars: formation --- 
brown dwarfs --- 
binaries: visual ---
stars: pre-main sequence}

\section{Introduction}
\label{sec:intro}

Most of the known planets outside of the solar system have been detected
through radial velocity measurements, and thus reside close to their parent
stars \citep[$a\lesssim5$~AU,][]{mar05,udr07}.
Meanwhile, a small number of companions with masses below 10~$M_{\rm Jup}$
have been detected at large separations through high-resolution imaging
\citep[$a\sim30$--300~AU,][]{cha04,laf08,kal08,mar08}.
The available data tend to favor core accretion as the mechanism responsible
for the close-in giant planets \citep{san04,fis05}, but gravitational
instability in disks and fragmentation of molecular cloud cores are
plausible (perhaps likely) alternatives for producing the wide companions
\citep{lod05,dob09}. The formation models for wide planetary-mass
companions are particularly in need of observational tests.

Companions are expected to form much more quickly through disk 
instability and cloud core fragmentation than core accretion 
\citep[$\tau\sim0.1$~Myr vs. $\tau\sim5$--10~Myr,][]{boss98,boss06,bur06,pol96}.
Thus, measurements of the frequency of planetary companions 
at very young ages can help identify the dominant mode for the formation
of these objects.
To provide data of this kind, we have performed a search for substellar
companions to 32 young brown dwarfs in the Taurus star-forming region
($\tau\sim1$~Myr, $d\sim140$~pc) using high-resolution images obtained
with the {\it Hubble Space Telescope}.  In this Letter, we present the
discovery of a companion that has a mass of 5--10~$M_{\rm Jup}$ 
and a projected separation of 15~AU from one of the brown dwarfs in this survey.

\section{Background}
\label{sec:bkgrnd}

The brown dwarf for which we have detected a substellar companion
is 2MASS J04414489+2301513 (henceforth 2M~J044144).
It is a member of the Taurus star-forming region and it has a spectral
type of M8.5 \citep{luh06tau}, which corresponds to a mass of 
$\sim20$~$M_{\rm Jup}$ for an age of 1 Myr \citep{cha00,luh03ic}.
2M~J044144 exhibits mid-IR excess emission and strong 
H$\alpha$ emission, indicating the presence of a circumstellar disk and 
active accretion, respectively \citep{luh06tau,luh10tau}.
\citet{kra07} identified a possible wide stellar companion
at a separation of $12\farcs4$ ($a=1700$~AU at 140~pc).
This candidate companion, 2MASS J04414565+2301580 (henceforth 2M~J044145),
has been spectroscopically confirmed as a young low-mass star, and hence a
likely member of Taurus \citep[M3-M4.5,][]{luh09tau,kra09}.

\section{Observations}

\subsection{WFPC2 Images}

We selected 2M~J044144 as one of the targets for a survey
of late-type members of Taurus with the Wide Field Planetary
Camera (WFPC2) on board the {\it Hubble Space Telescope}.
The targets were imaged through the F791W and F850LP filters with the PC
camera within WFPC2, which contained a $800\times800$ CCD
($0\farcs046$~pixel$^{-1}$).
In each filter, we obtained two images at each position in a two-point
diagonal dither pattern. For 2M~J044144, the exposure times of
the individual images were 260 and 160~s for F791W and F850LP, respectively.
The observations of 2M~J044144 were performed during
one orbit on 2008 August 20.

\begin{figure}
\plotone{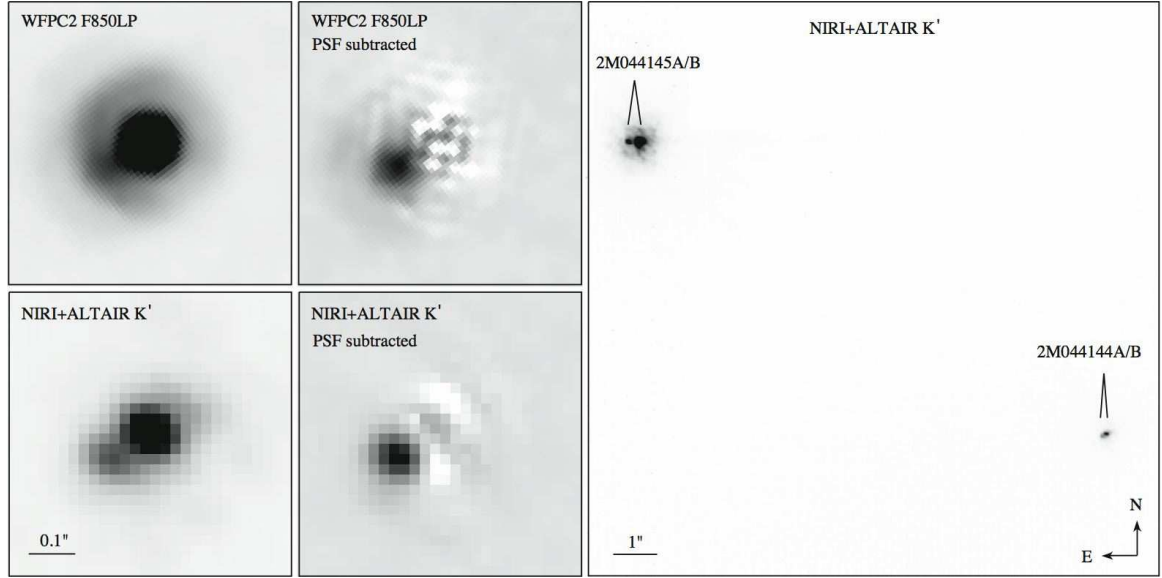}
\caption{
Left and middle: WFPC2 and NIRI+ALTAIR images of the young brown dwarf 
2M~J044144 before and after PSF subtraction ($0\farcs6\times0\farcs6$). 
The candidate companion has the same position relative to the brown
dwarf in each of these epochs, indicating that they have similar
proper motions (Table~\ref{tab:astro}). The separation of the companion
is $0\farcs105$, which corresponds to 15~AU at the distance of Taurus.
Right: An expanded version of the NIRI+ALTAIR image 
($13\arcsec\times13\arcsec$) showing both 2M~J044144
and the young low-mass star 2M~J044145.  A candidate substellar companion
to the latter is detected in these data ($\rho=0\farcs23$). The components
of 2M~J044144 and 2M~J044145 may comprise a young, low-mass quadruple system.
}
\label{fig:image}
\end{figure}

We reduced the WFPC2 images using the MultiDrizzle software \citep{koe02}.
To facilitate the detection of low-mass companions, we 
applied point spread function (PSF) subtraction to each brown dwarf in our
survey using techniques described by \citet{luh05wfpc}. This process
revealed a possible companion to 2M~J044144, which we refer to as 2M~J044144~B.
The images of this pair in F850LP before and after PSF subtraction are
shown in Figure~\ref{fig:image}.
We fit PSFs to both the primary and candidate companion to measure
their flux ratios and astrometric positions in the two filters.
In addition to PSF subtraction of the primaries, we identified all point
sources appearing in the WFPC2 images and measured
aperture photometry for them. In the case of 2M~J044144,
the PSF of the candidate companion was subtracted prior to the photometric
measurement of the primary. The photometry for companion was
computed by combining the aperture photometry of the primary with the flux
ratios produced by the PSF fits. The errors in the photometry for the primary
are dominated by the uncertainty in the correction for charge transfer
efficiency while the errors in the measurements for the companion
have a comparable contribution from the process of PSF fitting.
The astrometric offsets and photometry for the two objects are
presented in Tables~\ref{tab:astro} and \ref{tab:phot}, respectively.
Additional details concerning the observations and data analysis
will be provided in a forthcoming paper that presents the entire survey.

\subsection{NIRI+ALTAIR Images}

To constrain the relative proper motions of 2M~J044144
and its candidate companion, we obtained images of the pair at a second
epoch using NIRI \citep{hod03} in conjunction with
the ALTAIR adaptive optics system at the Gemini North telescope. 
NIRI was configured with the f/32 camera, which provided a 
plate scale of $0\farcs0214$~pixel$^{-1}$ and a field of view of
$22\arcsec\times22\arcsec$.
Because natural guide stars were unavailable, 
ALTAIR was operated in the laser guide star mode.
The tip/tilt correction was performed with 2M~J044145,
which is the possible wide companion to 2M~J044144 that
was described in \S~\ref{sec:bkgrnd}.
We imaged 2M~J044144 through the $H$ and $K\arcmin$ filters
using individual exposure times of 1 and 20~s.
For each exposure time and filter, one image was taken at each position
in a $3\times3$ dither pattern. 
After one set of dithered images was collected for all of the combinations
of exposure time and filter, the entire sequence of observations was repeated.
To provide a PSF for fitting the primary and candidate companion,
we also observed 2MASS 04383110+2310107. It was selected as the PSF star
because it was similar to the science target in its sky position and the
optical brightness, separation, and position angle of its tip/tilt star.

The individual images were divided by flat field exposures taken with the
Gemini Facility Calibration Unit.
The resulting images in a given dither sequence were registered and
combined,
producing two final images of 2M~J044144
for each combination of exposure time (1 and 20~s) and filter
($H$ and $K\arcmin$). All of the combined images of 2M~J044144
and the PSF star exhibited FWHM$\sim0\farcs09$, which is similar
to the image quality from WFPC2. 
To measure the relative positions and flux ratios of the components
of 2M~J044144, we performed PSF fitting 
with the four long exposures in the same manner as done for the WFPC2 data.
In Figure~\ref{fig:image}, we present one of the two long $K\arcmin$-band
exposures before and after PSF subtraction alongside the WFPC2 images.
We also show a wider area of this NIRI image that
encompasses the tip/tilt star, 2M~J044145, which is resolved
into a pair of objects that are separated by $0\farcs23$.
We applied PSF fitting to the components of this pair as well to
measure their flux ratios.  This was done with the short exposures
since the primary is saturated in the 20~s data.

Our measurements of the astrometric offsets between 2M~J044144~A and B
are listed in Table~\ref{tab:astro}.
We have adopted the average values produced by PSF fitting of the four long
exposures. We have derived photometry for the components of 2M~J044144 and
2M~J044145 by combining the flux ratios from the PSF fitting with the
unresolved $H$ and $K_s$ data from the Point Source Catalog of the Two-Micron
All-Sky Survey \citep[2MASS,][]{skr06}. We have assumed that the flux ratios
at $K_s$ are the equal to the ratios that we have measured at $K\arcmin$.

\section{Analysis}

\subsection{Evidence of Binarity}

\begin{figure}
\plotone{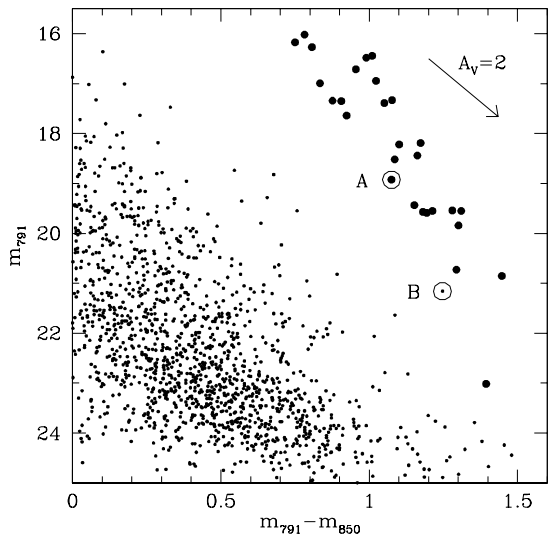}
\caption{
Color-magnitude diagram constructed from WFPC2 images of
32 young brown dwarfs in the Taurus star-forming region.
We show the targeted brown dwarfs that have unsaturated photometry
(large points) and all other point sources in these images (small points).
The components of 2M~J044144 are indicated (open circles,
Figure~\ref{fig:image}).
The secondary appears near the bottom of the sequence of known substellar
members of Taurus, which is consistent with a young low-mass companion.
}
\label{fig:iz}
\end{figure}

To investigate whether 2M~J044144~B is a companion to 2M~J044144~A,
we begin by plotting a color-magnitude diagram in Figure~\ref{fig:iz}
from photometry at F791W and F850LP for all unsaturated
sources in our entire WFPC2 survey of 32 brown dwarfs in Taurus.
Four of the substellar primaries are absent from the diagram because
they are saturated in one of the filters.
The faintest primary is 2MASS J04373705+2331080, which
is the coolest known member of Taurus \citep[L0,][]{luh09tau}.
As shown in Figure~\ref{fig:iz}, 2M~J044144~B appears
near the bottom of the sequence of the known members of Taurus, indicating
that it has the appropriate magnitudes and color for a low-mass companion.
Given the spectral types of the known members, these data suggest
a spectral type of M9.5-L0 for 2M~J044144~B.
In addition, the line connecting the primary and secondary is parallel to
the cluster sequence, which is consistent with the coevality that is
expected for the components of a binary system.

We can use the data in Figure~\ref{fig:iz} to
estimate the probability that our companion survey would detect a field star
that has the photometric properties of a low-mass Taurus member.
We assume that the sequence of Taurus members
becomes vertical at the faintest magnitudes in Figure~\ref{fig:iz}
($m_{791}>22$), which is supported by the location of the coolest known
member and the fact that the $I-Z$ colors of field dwarfs do not increase
significantly from M8 to mid-L \citep{sh00,dah02,dob02}.
In that case, $\sim7$ objects in Figure~\ref{fig:iz} (in addition to the
known brown dwarfs) have colors and magnitudes that are consistent with
membership in Taurus.
Therefore, the surface density of field stars that could be mistaken for
Taurus members is $\sim7$ divided by the total
area of the WFPC2 images, which is 160~arcmin$^2$.
The probability of a field star of this kind appearing within $0\farcs1$
of any of the 32 primaries in our survey is $\sim10^{-5}$. 

The two epochs of astrometry from WFPC2 and NIRI provide an additional
constraint on the companionship of 2M~J044144~B.
We first consider whether the astrometry for 2M~J044144~B is consistent
with the negligible motion expected for the vast majority of background stars.
Accurate measurements of the proper motion for the primary are not available.
Therefore, we adopt for the primary the average proper motion of the nearest
group of Taurus members \citep[$\mu_{\alpha}, \mu_{\delta}=+6.7$,
$-17.7$~mas~yr$^{-1}$,][]{luh09tau}.
Combining this proper motion with the astrometry from the first epoch,
we find that the separation and position angle of the candidate companion
would have become $0\farcs089$ and $111.5\arcdeg$ in the second epoch
if it had no proper motion, which is inconsistent with our measurements
in Table~\ref{tab:astro}. 

According to the data in Table~\ref{tab:astro}, 2M~J044144~A and B maintained
the same relative positions to within $\sim10$~mas over a period
of 1.15 years, indicating that their proper motions agree at a level of
$\sim9$~mas~yr$^{-1}$ in each direction. 
Only a small fraction of field stars are expected to
have proper motions that are similar to that of 2M~J044144~A.
For instance, using multi-epoch {\it Hubble} images of Chamaeleon~I
that are comparable in depth to our WFPC2 data in Taurus, we find
that $\sim0.3$\% of field stars have proper motions that are consistent with
those of cluster members at the level of the uncertainties in the
relative proper motions of 2M~J044144~A and B.
If we adopt this value for Taurus, which is similar to Chamaeleon~I
in its distance and the magnitude of its proper motion, then
the probability of a field star exhibiting WFPC2 photometry that is indicative
of a Taurus member, a separation of $\rho\leq0\farcs1$ from one of our 32
primaries, and a
proper motion similar to that of Taurus is $\sim3\times10^{-8}$.
Therefore, we conclude that 2M~J044144~B is a companion to 2M~J044144~A.

The possible wide companion 2M~J044145 appears within one
of the WFC arrays, but it is too heavily saturated for 
the detection of its candidate companion.
As a result, we cannot place constraints on the proper motions of
those two objects relative to each other.

\subsection{Physical Properties of Companion}

We can estimate the bolometric luminosity of 2M~J044144~B if
we apply a bolometric correction to its photometry at $K_s$.
In previous studies of young brown dwarfs, we have adopted bolometric
corrections derived for field dwarfs, but doing so may not be perfectly
valid for 2M~J044144~B since some of the IR colors of young objects later
than M9 differ from those of dwarfs \citep{cru09,luh09tau}.
To determine an appropriate value of BC$_K$ for 2M~J044144~B, we have estimated
bolometric luminosities for young brown dwarfs from M9.5--L0 for which
relatively complete spectral energy distributions have been measured,
consisting of the Taurus member
KPNO~4 \citep[M9.5,][]{bri02} and the young field dwarfs
2MASS J01415823$-$4633574 and 2MASS J02411151$-$0326587
\citep[L0,][]{kir06,cru09}.
We have excluded from this sample brown dwarfs that exhibit disk emission
in their mid-IR data or that have extinctions of $A_V>1$.
For each object, we constructed a spectral energy distribution using a
0.8-2.5~\micron\ spectrum from the NASA
Infrared Telescope Facility that was flux calibrated with 2MASS photometry,
3.2-9.2~\micron\ photometry from the {\it Spitzer Space Telescope}
\citep{luh09tau,luh10tau}, a linear interpolation of the fluxes between 
2.5~\micron\ and 3.2~\micron, and a Rayleigh-Jeans distribution longward of
9.2~\micron. By integrating the flux in each distribution and combining
the resulting luminosity with the $K_s$ photometry, we arrive at
BC$_K=3.42$, 3.39, and 3.41 for KPNO~4, 2MASS J01415823$-$4633574,
and 2MASS J02411151$-$0326587, respectively. These bolometric corrections
are larger than those of dwarfs by $\sim0.2$~mag \citep{gol04}.
Based on these measurements, we adopt BC$_K=3.4$ for 2M~J044144~B.

The $J-H$ color from 2MASS and the 0.8-1.8~\micron\ spectrum from
\citet{luh06tau} for 2M~J044144~A+B indicate $A_V\sim0$ when they are
compared to average intrinsic colors and spectra of young brown
dwarfs \citep{luh10tau}. In the $K$ band, the spectroscopy and 2MASS
photometry of the system and our resolved photometry for the primary
are slightly redder than a stellar photosphere with the spectral type of
the primary, which is probably caused by emission from a circumstellar
disk based on the excess emission detected at longer wavelengths 
\citep{luh10tau}.
Therefore, we have not applied an extinction correction to the
photometry for 2M~J044144~B. When we combine $K_s=14.94$, BC$_K=3.4$,
$M_{\rm bol \odot}=4.75$, and the average distance of Taurus
\citep[$d=140$~pc,][]{wic98,tor09}, we derive log~$L_{\rm bol}=-3.14$ for
2M~J044144~B.

\begin{figure}
\plotone{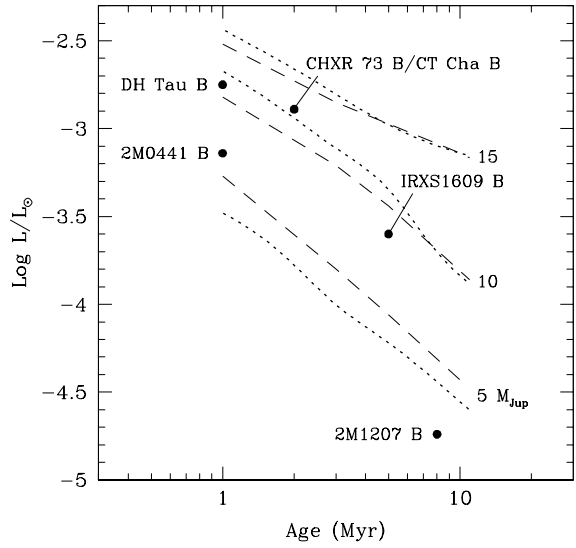}
\caption{
Luminosity estimates of 2M~J044144~B and other young low-mass companions
($M\lesssim15$~$M_{\rm Jup}$) compared to the luminosities as a function
of age predicted by the theoretical evolutionary models of
\citet[][dashed lines]{cha00} and \citet[][dotted lines]{bur97}.
CHXR~73~B and CT~Cha~B have nearly identical positions in this diagram.
These models indicate a mass of 5--10~$M_{\rm Jup}$ for 2M~J044144~B.
The luminosities of all of these companions have uncertainties near
$\pm0.12$~dex.
}
\label{fig:lbol}
\end{figure}

In Figure~\ref{fig:lbol}, we estimate a mass from the luminosity of
2M~J044144~B by comparing it to the values
predicted by theoretical evolutionary models for an age of 1~Myr, which
is the average age for members of Taurus and the isochronal age of the primary
\citep{luh09tau}. Given the uncertainties in its age (few Myr) and luminosity
($\pm0.12$~dex), 2M~J044144~B could have a mass between 5 and 10~$M_{\rm Jup}$
according to the models of \citet{cha00} and \citet{bur97}. 
For comparison, we include in Figure~\ref{fig:lbol} the luminosities of other
resolved companions that appear to have masses below $\sim15$~$M_{\rm Jup}$ and
that are younger than 10~Myr, which consist of 2M1207-3932~B \citep{cha04},
DH~Tau~B \citep{ito05}, CHXR~73~B \citep{luh06bin}, CT~Cha~B \citep{sch08},
and IRXS1609~B \citep{laf08}.

\section{Discussion}

The first planetary-mass companion to be detected through direct imaging
was 2MASS J12073346-3932539~B
\cite[$a=40$~AU, $M\sim4$~$M_{\rm Jup}$,][]{cha04,bil07,giz07,duc08},
which is bound to a young brown dwarf in the TW Hya association
\citep[$\tau\sim8$~Myr, $M\sim25$~$M_{\rm Jup}$,][]{giz02}.
The new companion that we have presented, 2M~J044144~B, is a younger
analog of 2MASS J12073346-3932539~B, thus demonstrating that
wide planetary-mass companions to brown dwarfs can form on a timescale
of $\tau\lesssim1$~Myr. This result is consistent with the rapid formation
expected from both gravitational instability in disks and fragmentation
of cloud cores. The latter is more likely to have produced these
two secondaries given their relatively large masses compared to the
primaries \citep{lod05}. 
Indeed, if 2M~J044144~A and B are members of a quadruple system,
then its hierarchical configuration further suggests
that 2M~J044144~B formed by cloud core fragmentation.
Additional data are needed to determine more definitively whether
2M~J044144~A/B and 2M~J044145~A/B comprise a quadruple system.

\acknowledgements

We acknowledge support from grant GO-11203 from the Space Telescope Science
Institute, grant AST-0544588 from the National Science Foundation (K. T.,
K. L.), and Theresa Mall Mullarkey (K. M.). We thank Allison Youngblood,
Steven Mohammed, Ijeoma Ekeh, and Jaclyn Payne for assistance with the data
analysis. The Gemini data were obtained
through program GN-2009B-DD-3. Gemini Observatory is operated
by AURA under a cooperative agreement with the NSF on behalf of the
Gemini partnership: the NSF (United States), the Particle Physics and
Astronomy Research Council (United Kingdom), the National Research
Council (Canada), CONICYT (Chile), the Australian Research Council
(Australia), CNPq (Brazil) and CONICET (Argentina).
The Center for Exoplanets and Habitable Worlds is supported by the
Pennsylvania State University, the Eberly College of Science, and the
Pennsylvania Space Grant Consortium. 
MultiDrizzle is a product of the 
Space Telescope Science Institute, which is operated by AURA for NASA.

\clearpage

\begin{deluxetable}{llll}
\tabletypesize{\scriptsize}
\tablewidth{0pt}
\tablecaption{Astrometry for 2MASS J04414489+2301513 B\label{tab:astro}}
\tablehead{
\colhead{Instrument} &
\colhead{$\rho$} &
\colhead{P.A.} &
\colhead{Date} \\
\colhead{} &
\colhead{(arcsec)} &
\colhead{(deg)} &
\colhead{}}
\startdata
WFPC2       & 0.105$\pm$0.004 & 120.4$\pm$2.2 & 2008 Aug 20 \\
NIRI+ALTAIR & 0.108$\pm$0.005 & 121.0$\pm$2.6 & 2009 Oct 13 \\
\enddata
\end{deluxetable}

\begin{deluxetable}{lllll}
\tabletypesize{\scriptsize}
\tablewidth{0pt}
\tablecaption{Photometry for 2MASS J04414489+2301513 and
2MASS J04414565+2301580\label{tab:phot}}
\tablehead{
\colhead{2MASS} &
\colhead{m$_{791}$} &
\colhead{m$_{850}$} &
\colhead{$H$} &
\colhead{$K_s$} 
}
\startdata
J04414489+2301513 A & 18.93~$\pm$~0.05 & 17.85~$\pm$~0.05 & 13.94~$\pm$~0.03 & 13.40~$\pm$~0.03 \\
J04414489+2301513 B & 21.16~$\pm$~0.10 & 19.91~$\pm$~0.10 & 15.62~$\pm$~0.10 & 14.94~$\pm$~0.10 \\
J04414565+2301580 A & \nodata & \nodata & 10.17~$\pm$~0.02 &  9.94~$\pm$~0.02 \\
J04414565+2301580 B & \nodata & \nodata & 13.03~$\pm$~0.06 & 12.59~$\pm$~0.06 \\
\enddata
\end{deluxetable}

\clearpage

\end{document}